\documentclass[runningheads]{cl2emult}
\usepackage{amssymb}
\begin{document}

% DEFINITIONS
%%%%%%%%%%%%%%%%%%%%%%%%%%%%%%%%%%%%%%%%%%%%%%%%%%
\def\C{{\cal C}}
\def\H{{\cal H}}
\def\O{{\cal O}}
\def\A{{\cal A}}
\def\B{{\cal B}}
\def\S{{\cal S}}
\def\PH{{\cal PH}}
\def\shalf{\hbox{${\textstyle{1\over 2}}$}}
\def\op#1{\hbox{\sf #1}}
\def\ihbar{\hbox{${\textstyle{i\over\hbar}}$}}
\def\Mhbar{\hbox{${\textstyle{M\over\hbar}}$}}
\def\hbarM{\hbox{${\textstyle{\hbar\over M}}$}}
\def\reals{\hbox{${\mathbb R}$}}
\def\complex{\hbox{${\mathbb C}$}}
\def\b#1{\bar{#1}} 
\def\t#1{\tilde{#1}}
%%%%%%%%%%%%%%%%%%%%%%%%%%%%%%%%%%%%%%%%%%%%%%%%%%

\title*{Decoherence, a Dynamical Approach to Superselection Rules ?}
\toctitle{Decoherence, a Dynamical Approach to Superselection Rules ?}
\titlerunning{Decoherence and Superselection}

\author{Domenico Giulini\inst{}}
\authorrunning{Domenico Giulini}
\institute{Theoretische Physik, Universit\"at Z\"urich,\\
           Winterthurerstrasse~190, CH-8057 Z\"urich, 
           Switzerland}

\maketitle

\begin{abstract}
It is well known that the dynamical mechanism of decoherence
may cause apparent superselection rules, like that of molecular 
chirality. These `environment-induced' or `soft' 
superselection rules may be contrasted with `hard' 
superselection rules, like that of electric charge, whose existence 
is usually rigorously demonstrated by means of certain symmetry 
principles. We address the question of whether this distinction 
between `hard' and `soft' is well founded and argue that, 
despite first appearance, it might not be. For this we first review 
in detail some of the basic structural properties of the spaces 
of states and observables in order to establish a fairly precise 
notion of superselection rules. We then discuss two examples: 
1.)~the Bargmann superselection rule for overall mass in ordinary 
quantum mechanics, and 2.)~the superselection rule for charge in 
quantum electrodynamics. 
\end{abstract}

\section{Introduction}
To explain the (apparent) absence of interferences between
macroscopically interpretable states -- like states describing
spatially localized objects -- is the central task for any attempt
to resolve the measurement problem. First attempts in this direction
just imposed additional rules, like that of the Copenhagen school,
who {\it defined} a measurement device as a system whose state-space
is classical, in the sense that the superposition principle is fully
broken: superpositions between any two states simply do not exist.
In a more modern language this may be expressed by saying that
any two states of such a system are {\it disjoint}, i.e., separated
by a superselection rule (see below). Proper quantum mechanical
systems, which in isolation do obey the superposition principle, 
can then inherit superselection rules when coupled to such classical 
measurement devices.

Whereas there can be no doubt that the notion of classicality, as 
we understand it here, is mathematically appropriately encoded in 
the notions of {\it disjointness} and {\it superselection rules}, 
there still remains the physical question how these structures 
come to be imposed. In particular, if one believes that fundamentally 
all matter is described by some quantum theory, there is no room 
for an independent classical world. Classicality should be a feature
that is emerging in accordance with, and not in violation of, the 
basic rules of quantum mechanics.  This is the initial credo of those 
who believe in the program of {\it decoherence} \cite{GJKKSZ},  which
aims to explain classicality by means of taking into account 
dynamical interactions with ambient systems, like the ubiquitous 
natural environment of the situation in question. (Note: It is not 
claimed to resolve the full measurement problem.)
This leads to the notion of `environment-induced superselection 
rules'~\cite{Zurek}.
 
Fundamental to the concept of dynamical decoherence is the notion of 
`delocalization'~\cite{Joos}. The intuitive idea
behind this is that through some dynamical process certain state 
characteristics (`phase relations'), which were locally accessible at 
one time, cease to be locally accessible in the course of the dynamical
evolution. Hence locally certain superpositions cannot be 
verified anymore and an apparent obstruction to the superposition 
principle results. Such mechanisms are  considered responsible for 
the above mentioned environment-induced superselection rule, of which 
a famous physical example is that of molecular 
chirality~(see e.g. \cite{Wightman1} and references therein). 
It has been established in many calculations of realistic
situations that such dynamical processes of delocalization can be  
extremely effective over short time scales. But it is also 
intuitively clear that, mathematically speaking, it will never be 
strict in any finite time. Hence one will have to deal with notions  
of approximate- respectively asymptotic (for $t\rightarrow\infty$) 
superselection rules and disjointness of 
states~\cite{Primas,Kupsch}, which needs some mathematical care.

Since for finite times such dynamical superselection rules 
are only approximately valid, they are sometimes called `soft'. 
In contrast, `hard' superselection rules are those which are 
rigorously established mathematical results within the 
kinematical framework of the theory, usually based on symmetry
principles (see section~3 below), or on first principles of local 
QFT, like in the proof for the superselection rule for
electric charge~\cite{Strocchi-Wightman}. Such presentations 
seem to suggest that there is no room left for a dynamical
origin of `hard' superselection rules, and that hence these
two notions of superselection rules are really distinct. 
However, we wish to argue that at least {\it some} of the 
existing proofs for `hard' superselection rules give a false 
impression, and that quite to the contrary they actually 
need some dynamical input in order to be physically convincing.
We will look at the case of Bargmann's superselection rule for total
mass in ordinary quantum mechanics (which is clearly more of an 
academic example) and that of charge in QED. The discussion of the 
latter will be heuristic insofar as we will pretend that QED is 
nothing but quantum mechanics (in the Schr\"odinger representation) 
of the infinite-dimensional (constrained) Hamiltonian system given 
by classical electrodynamics. For a brief but general orientation 
on the subject of superselection rules and the relevant references we
refer to Wightman's survey~\cite{Wightman2}. 

Let us stress again that crucial to the ideas presented here is of 
course that `delocalized' does not at all mean `destroyed', and that
hence the loss of quantum coherence is only an {\it apparent} one. 
This distinction might be considered irrelevant FAPP (for all 
practical purposes) but it is important in attempts to understand 
apparent losses of quantum coherence {\it within} the standard 
dynamical framework of quantum mechanics.

As used here, the term `local' usually refers to locality in 
the (classical) configuration space $Q$ of the system, where we 
think of quantum states in the Schr\"odinger representation, i.e., as 
$L^2$-functions on $Q$. Every parametrization of $Q$ then defines
a partition into `degrees of freedom'. Locality in $Q$ is a more general
concept than locality in ordinary physical space, although the latter
forms a particular and physically important special case. Moreover, on a
slightly more abstract level, one realizes that the most general
description of why decoherence appears to occur is that only a
{\it restricted} set of so-called physical observables are at ones
disposal, and that  {\it with respect to those} the relevant `phase
relations' {\it apparently} fade out of existence. It is sometimes
convenient to express this by saying that decoherence occurs only
with respect (or relative) to a `choice' of observables~\cite{Landsman}.
Clearly this `choice' is not meant to be completely free, since it has
to be compatible with the dynamical laws and the physically realizable
couplings (compare~\cite{Joos}). (In this respect the situation bears 
certain similarities to that of `relevant' and `irrelevant' degrees 
of freedom in statistical mechanics.) But to fully control those is 
a formidable task -- to put it mildly. In any case it will be necessary
to assume some a priori characterizations of what mathematical objects 
correspond to observables, and to do this in such a general fashion 
that one can effectively include superselection rules. This will be 
done in the  next section.

\section{Elementary Concepts}
In this section we wish to convey a feeling for some
of the concepts underlying the notion of superselection rules.
We will take some care and time to do this, since many misconceptions
can (and do!) arise from careless uses of these concepts.
To gain intuition it is sometimes useful to dispense with some 
technicalities associated with infinite dimensions and 
continuous spectra and just look at finite dimensional situations; 
we will follow this strategy where indicated. We use the following,
generally valid notations: $\H$ denotes a Hilbert space, $B(\H)$
the algebra of bounded operators on $\H$. The antilinear
operation of taking the hermitean conjugate is denoted by
$*$ (rather than $\dagger$), which makes $B(\H)$ a $*$-algebra.
Given a set $\{A_{\lambda}\}$ where $\lambda\in\Lambda$ (= some
index set), then by $\{A_{\lambda}\}'$ we denote the `commutant'
of $\{A_{\lambda}\}$ in $B(\H)$, defined by
\begin{equation}
\{A_{\lambda}\}':=\{B\in B(\H)\,\vert\, BA_{\lambda}=A_{\lambda}B,\
\forall \lambda\in\Lambda\}.
\label{commutant}
\end{equation}
Note that if the set $\{A_{\lambda}\}$ is left invariant under 
the $*$-map (in this case we call the set `self-adjoint'), then 
$\{A_{\lambda}\}'$ is a $*$-subalgebra of $B(\H)$. Also, the 
definition (\ref{commutant}) immediately implies that 
\begin{equation}
\A\subseteq\B\Rightarrow \B'\subseteq\A'.
\label{commutant-inclusion}
\end{equation}

\subsection{Superselection Rules}
There are many different ways to give a structural definition of 
superselection rules. Some stress the notion of {\it state} others 
the notion of {\it observable}. Often this dichotomy seems to 
result in very different attitudes towards the fundamental 
significance of superselection rules. This really seems 
artificial in a quantum mechanical context. In quantum 
{\it field} theory, i.e., if the underlying classical system has 
infinitely many degrees of freedom, the situation seems more 
asymmetric. This is partly due to the mathematical difficulties 
to define the full analog of the Schr\"odinger representation, i.e.,
to just construct the Hilbert space of states as $L^2$ space over 
the classical configuration space. In this paper we will partly 
ignore this mathematical difficulty and proceed heuristically by 
assuming that such a Schr\"odinger representation (of QED) 
exists to some level of rigour.

In traditional quantum mechanics, which stresses the notion of 
state, a system is fundamentally characterized by a Hilbert 
space, $\H$, the vectors of which represent (pure) states. We 
say `represent' because this labeling by states through vectors 
is redundant: non-zero vectors which differ by an overall complex 
number label the same state, so that states can also be labeled 
by rays.  We will use $\PH$ to denote the space of rays in $\H$.
In many cases of interest this Hilbert space is of course just 
identified with the space of $L^2$-functions over the classical 
configuration space. Now, following the original definition given 
by $W^3$~\cite{WWW1}, we say that a superselection rule operates on
$\H$, if not all rays represent pure states, but only those which 
lie entirely in certain mutually orthogonal subspaces 
$\H_i\subset\H$, where  
\begin{equation}
\H=\bigoplus_i\H_i.
\label{ssrsum} 
\end{equation}
The only rays which correspond to pure states are those in the 
disjoint union 
\begin{equation}
\bigcup_i\PH_i\,.
\label{ssrunion}
\end{equation}
Since no vector which lies skew to the partition (\ref{ssrsum})
can, by assumption, represent a pure state, the superposition 
principle must be restricted to the $\H_i$. Moreover, since 
observables map pure states to pure states, they must leave the 
$\H_i$ invariant and hence all matrix-elements of observables 
between vectors from different sectors vanish. 
The $\H_i$ are called {\it coherent sectors} if the observables act 
irreducibly on them, i.e., if no further decomposition is possible; 
this is usually implied if a decomposition (\ref{ssrsum}) is 
written down. States which lie in different coherent sectors are 
called {\it disjoint}. Note that disjointness of states is essentially 
also a statement about observables, since it means orthogonality of 
the original states {\it and} the respective states created from 
those with {\it all} observables. The existence of disjoint states 
is the  characteristic feature of superselection rules.

From this we see that a partition (\ref{ssrsum}) into coherent sectors 
implies that the set of physical observables is strictly smaller than 
the set of all self-adjoint (w.l.o.g. bounded) operators on $\H$. 
It can be characterized by saying that observables are those self 
adjoint operators on $\H$ which commute with the orthogonal projectors 
$P_i:\H\rightarrow\H_i$. So the $P_i$ are themselves observables and 
generate the center (see below) of the algebra of observables.

This suggests a `dual', more algebraic way to look at superselection 
rules, which starts with the algebra of observables $\O$. Then
superselection rules are said to occur if the algebra of 
observables, $\O$, -- which we think of as being given by 
bounded operators on some Hilbert space
%% BEGIN FOOTNOTE
$\H$\footnote{In Algebraic Quantum Mechanics one associates
to each quantum system an abstract $C^*$-algebra, $\C$, which
is thought of as being the mathematical object that fully
characterizes the system in isolation, i.e. its intrinsic or
`ontic' properties. But this is not yet what we call the algebra
of observables. This latter algebra is not uniquely determined by
the former. It is obtained by studying faithful representations of
$\C$ in some Hilbert space $\H$, such that $\C$ can be identified
with some subalgebra of $B(\H)$ (the bounded operators on $\H$).
This is usually done by choosing a reference state (positive
linear functional) on $\C$ and performing the GNS construction.
Then $\C$ inherits a norm which is used to close $\C$ (as 
topological space) in $B(\H)$. It is this resulting algebra
which corresponds to our $\O$. Technically speaking it is a
von~Neumann algebra which properly contains an embedded copy of $\C$. 
The added observables (those in $\O-\C$) do not describe intrinsic but
{\it contextual} properties. For example, it may happen that $\O$ has
non-trivial center whereas $\C$ doesn't. In this case the superselection
rules described by $\O$ are contextual. See \cite{Primas} for a more 
extended discussion of this point.}
%% END OF FOOTNOTE
 -- has a non-trivial center $\O^c$. Recall that
\begin{equation}
\O^c:=\{A\in\O\,\vert\,AB=BA,\ \forall B\in\O\}.
\label{center}
\end{equation}
Suppose $\O^c$ is generated by self-adjoint elements 
$\{C_{\mu},\mu=1,2,..\}$ which have simultaneous eigenspaces 
$\H_i$, then the $\H_i$'s are just the coherent sectors. Indeed, as
already remarked,  
matrix elements of operators from $\O$ between states from different 
coherent sectors (i.e. differing in the eigenvalue of at least one 
$C_{\mu}$) necessarily vanish. Thus if $\phi_1$ and $\phi_2$ 
are two non-zero vectors from $\H_i$ and $\H_j$ with $i\not =j$, their 
superposition $\phi:=\phi_i+\phi_2$ defines a state whose density matrix 
$\rho:=P_{\phi}$ (=orthogonal projector onto the ray generated by 
$\phi$) satisfies
\begin{equation}
\hbox{tr}(\rho A)=\hbox{tr}((\lambda_1\rho_1+\lambda_2\rho_2)A),
\quad\forall A\in\O,
\label{convcomb}
\end{equation}
where $\lambda_{1,2}=\Vert\phi_{1,2}\Vert^2/\Vert\phi\Vert^2$ and
$\rho_{1,2}=P_{\phi_{1,2}}$. This means that $\rho$ is a non-pure 
state of $\O$, since it can be written as a non-trivial convex 
combination of other density matrices. Hence we come back to the 
statements expressed by (\ref{ssrsum}) and (\ref{ssrunion}).
Also note the following: in quantum mechanics the decomposition of a 
non-pure density matrix as a convex combination of pure density 
matrices -- the so-called extremal decomposition -- is generically not 
unique, thus preventing the (ignorance-) interpretation as statistical 
%% BEGIN FOOTNOTE
``mixtures''.\footnote{Hence the term `mixture' for a non-pure state  
is misleading since we cannot tell the components and hence have no 
ensemble interpretation. For this reason we will say `non-pure state' 
rather than `mixture'.} 
%% END FOOTNOTE
However, for the special density matrices of the form 
$\rho=\vert\phi\rangle\langle\phi\vert$, where $\vert\phi\rangle\in\H$, 
the extremal decomposition {\it is} unique and given by  
$\phi=\sum_i\lambda_iP_{\phi_i}$, where $\phi_i$ is the orthogonal 
projection of $\phi$ into $\H_i$, $P_{\phi_i}$ the orthogonal 
projector onto $\phi_i$'s ray, and 
$\lambda_i=\Vert\phi_i\Vert^2/\Vert\phi\Vert^2$. This is the 
relevance of superselection rules for the measurement problem: 
to produce {\it unique} extremal decompositions -- and hence 
statistical `mixtures' in the proper sense of the word -- into 
an ensemble of pure states. There is a long list of papers dealing 
with the mathematical problem of how superselection sectors can 
arise dynamically; see e.g. \cite{Hepp,Pfeifer,Araki,Kupsch} and 
the more general discussions in \cite{Landsman,Primas}.

\subsection{Dirac's Requirement}
Dirac was the first who spelled out certain rules concerning the 
spaces of states and observables~\cite{Dirac}. He defined the 
notion of {\it compatible} (i.e., simultaneously performable) 
{\it observations}, which mathematically are represented by a set of 
commuting observables, and the notion of a {\it complete set}    
of such observables, which is meant to say that there is precisely 
one state for each set of simultaneous ``eigenvalues''. 
Starting from the hypothesis that states are faithfully represented 
by rays, Dirac deduced that a complete set of such mutually compatible 
observables existed. But this only makes sense if all the 
observables in question have purely discrete spectra. 

In the general case one has to proceed differently: We heuristically 
define Dirac's requirement as the statement, that {\it there exists 
at least one complete set of mutually compatible observables} and show  
how it can be rephrased mathematically so that it applies to all cases. 
In doing this we essentially follow Jauch's exposition~\cite{Jauch1}. 
To develop a feeling for what is involved, we will first describe some 
of the consequences of Dirac's requirement in the most simplest case: 
a finite dimensional Hilbert space. We will use this insight to rephrase 
it in such a way to stay generally valid in infinite dimensions.

\subsubsection{Gaining intuition in finite dimensions.}
So let $\H$ be an $n$-dimensional complex Hilbert-space,
then $B(\H)$ is the algebra of complex $n\times n$ matrices.
Physical observables are represented by hermitean matrices in
$B(\H)$, but we will explicitly {\it not} assume the converse,
namely that {\it all} hermitean matrices correspond to physical
observables. Rather we assume that the physical observables are
somehow given to us by some set $\S$ of hermitean matrices.
This set does not form an algebra, since taking products and
complex linear combinations does not preserve hermiticity.
But for mathematical reasons it would be convenient to have
such an algebraic structure, and just work with the algebra
$\O$ generated by this set, called the {\it algebra of observables}.
[Note the usual abuse of language, since only the hermitean
elements in $\O$ are observables.]
But for this replacement of $\S$ by $\O$ to be allowed $\S$ must
have been a set of hermitean matrices which is uniquely determined
by $\O$, for otherwise we can not reconstruct the set $\S$ from
$\O$. To grant us this mathematical convenience we assume that
$\S$ was already maximal, i.e. that $\S$ already contains all the
hermitean matrices that it generates. But we stress that there
seems to be no obvious reason why in a particular practical situation 
the set of physically realizable observables should be maximal in 
this sense.

We may choose a set $\{O_1,\dots O_m\}$ of hermitean generators
of $\O$. Then $\O$ may be thought of as the set of all
complex polynomials in these (generally non-commuting) matrices.
But note that we need not consider higher powers than $(n-1)$ of
each $O_i$, since each complex $n\times n$ matrix $O$ is a zero
of its own characteristic polynomial $p_O$, i.e. satisfies $p_O(O)=0$,
by the theorem of Cayley-Hamilton. Since this polynomial is
of order $n$, $O^n$ can be re-expressed by a polynomial in $O$
of order at most $(n-1)$. For example, the $*$-algebra generated
by a single hermitean matrix $O$ can be identified with the set
of all polynomials of degree at most $(n-1)$ and whose
multiplication law is as usual, followed by the procedure of
reducing all powers $n$ and higher of $O$ via $p_O(O)=0$.

Now let $\{A_1,\cdots,A_m\}=:\{A_i\}$ be a complete set of mutually 
commuting observables. It is not difficult to show that there 
exists an observable $A$ and polynomials $p_i$, $i=1,\cdots,m$ 
such that $A_i=p_i(A)$ (see \cite{Isham} for a simple proof). This
actually 
means that the algebra generated by $\{A_i\}$ is just the $n$-dimensional 
algebra of polynomials of degree at most $n-1$ in $A$ (see below for 
justification), which we call $\A$. This algebra is abelian, which 
is equivalently expressed by saying that $\A$ is contained in its 
commutant (compare~(\ref{commutant})):
\begin{equation}
\A\subseteq\A' \qquad\fbox{\hbox{`$\A$ is abelian'}}
\label{abelian}
\end{equation}

Now comes the requirement of completeness. In terms of $A$ it is 
easy to see that it is equivalent to the condition that $A$ 
has a simple spectrum (i.e. the eigenvalues are pairwise distinct). 
This has the following consequence: Let $B$ be an observable that 
commutes with $A$, then $B$ is also a function of $A$, i.e., 
$p_B(B)=A$ for some polynomial $p_B$. The proof is simple: We 
simultaneously diagonalize $A$ and $B$ with eigenvalues $\alpha_a$
and $\beta_a$, $a=1,\cdots,n$. We wish to find a polynomial of 
degree $n-1$ such that $p_B(\alpha_a)=\beta_a$. Writing 
$p_B(x)=a_{n-1}x^{n-1}+\cdots+a_0$, this leads to a system of 
$n$ linear equations ($\alpha_a^b:=$ $b^{\rm th}$ power of $\alpha_a$) 
\begin{equation}
\sum_{b=0}^{n-1}\alpha_a^ba_b=\beta_a,\qquad\hbox{for}\ a=1,\cdots,n
\label{ABequation}
\end{equation}
for the $n$ unknowns $(a_0,\cdots,a_{n-1})$. Its determinant is of 
course just the Vandermonde determinant for the $n$ tuple 
$(\alpha_1,\cdots,\alpha_n)$:
\begin{equation}
\hbox{det}\{\alpha_a^b\}=\prod_{a<b}(\alpha_a-\alpha_b),
\label{vandermonde}
\end{equation}
which is non-zero if and only if (=iff) $A$'s spectrum is simple. 
This implies that every observable that commutes with $\A$ is 
already contained in $\A$. (It follows from this that the algebra 
generated by $\{A_i\}$ is equal to, and not just a subalgebra of, 
the algebra generated by $\{A\}$, as stated above.) 
Since a $*$-algebra is generated by its self-adjoint elements 
(observables), $\A$ cannot be properly enlarged as abelian $*$-algebra 
by adding more commuting generators. In other words, $\A$ is {\it maximal}. 
Since $\A'$ is a $*$-algebra, this can be equivalently expressed by 
\begin{equation}
\A'\subseteq \A \qquad\fbox{\hbox{`$\A$ is maximal'}}
\label{maximal}
\end{equation}

Equations (\ref{abelian}) and (\ref{maximal}) together are equivalent
to Dirac's condition, which can now be stated in the following 
form, first given by Jauch~\cite{Jauch1}: the algebra of 
observables $\O$ contains a maximal 
abelian $*$-subalgebra $\A\subseteq\O$, i.e.,
\begin{equation}
\fbox{$\hbox{Dirac's requirement, $1^{\rm st}$ version:}\ 
\exists\ \A\subseteq\O\ \hbox{satisfying}\ \A=\A'$}
\label{Dirac1}
\end{equation}
 
This may seem as if Dirac's requirement could be expressed in 
purely algebraic terms. But this is deceptive, since the very notion 
of `commutant' (compare (\ref{commutant})) makes reference to the 
Hilbert space $\H$ through $B(\H)$. Without further qualification 
the term `maximal' always means maximal {\it in} 
$B(\H)$.\footnote{\label{maximality}The condition for an abelian
$\A\subseteq\O$ to be maximal in $\O$ would be $\A=\A'\cap\O$. 
Such abelian subalgebras {\it always} exist (use Zorn's Lemma 
to show this), in contrast to those $\A\subseteq\O$ which 
satisfy the stronger condition to be maximal in the ambient 
algebra $B(\H)$, which need not exist for a given 
$\O\subset B(\H)$.}

This reference to $\H$ can be further clarified by yet another 
equivalent statement of Dirac's requirement. Since $\A$ consists 
of polynomials in the observable $A$, which has a simple spectrum, 
the following is true: there exists a vector $\vert g\rangle\in\H$, 
such that for {\it any} vector $\phi\in\H$ there exists a polynomial 
$p_{\phi}$ such that 
\begin{equation}
p_{\phi}(A)\vert g\rangle=\vert\phi\rangle.
\label{cyclic1}
\end{equation}
Such a vector $\vert g\rangle$ is called a generating or 
{\it cyclic vector} for $\A$ in $\H$. The proof is again very simple: 
let $\{\phi_1,\cdots,\phi_n\}$ be the pairwise distinct, non-zero 
eigenvectors of $A$ (with any normalization); then choose
\begin{equation}
\vert g\rangle=\sum_{i=1}^{n}\vert\phi_i\rangle.
\label{cyclic2}
\end{equation}
Equation (\ref{cyclic1}) now defines again a system of $n$ linear 
equations for the $n$ coefficients $a_{n-1},\cdots,a_0$ of the polynomial 
$p_{\phi}$, whose determinant is again the Vandermonde determinant 
(\ref{vandermonde}) for the $n$ eigenvalues $\alpha_1,\cdots,\alpha_n$
of $A$. Conversely, if $A$ had an eigenvalue, say $\alpha_1$, with 
eigenspace $\H_1$ of two or higher dimensions, then such a cyclic 
$\vert g\rangle$ cannot exist. To see this, suppose it did, and let 
$\vert\phi_1^{\perp}\rangle\in\H_1$ be orthogonal to the projection of 
$\vert g\rangle$ into $\H_1$. 
Then $\langle\phi_1^{\perp}\vert p(A)g\rangle=0$ for all polynomials $p$.
Thus $\vert\phi_1^{\perp}\rangle$ is unreachable, contradicting our 
initial assumption. Hence a simple spectrum of $A$ is equivalent to the 
existence of a cyclic vector. 
      
\subsubsection{The general case.}
In infinite dimensions we have to care a little more about the 
topology on the space of observables, since here there are many 
inequivalent ways to generalize the finite dimensional case.
The natural choice is the so-called `weak topology', which is 
characterized by declaring that a sequence $\{A_i\}$ of observables 
converges to the observable $A$ if the sequence $\langle\phi\vert
A_i\vert\psi\rangle$ of complex numbers converges to
$\langle\phi\vert\A\vert\psi\rangle$ for all
$\vert\phi\rangle,\vert\psi\rangle\in\H$. Hence one also requires 
that the algebra of observables is weakly closed (i.e., closed 
in the weak topology). Such a weakly closed $*$-subalgebra of
$B(\H)$ is called a $W^*$- or von-Neumann-algebra (we shall 
use the first name for brevity).

A crucial and extremely convenient point is, that the weak 
topology is fully encoded in the operation of taking 
the commutant (see (\ref{commutant})), in the following sense: 
Let $\{A_{\lambda}\}$ be any subset of $B(\H)$, then $\{A_{\lambda}\}'$ 
is automatically weakly closed (see \cite{Jauch1} p~716 for a 
simple proof) and hence a $W^*$-algebra. Moreover, the weak closure 
of a $*$-algebra $\A\subseteq B(\H)$ is just given by $\A''$ 
(the commutant of the commutant). Hence we can characterize a 
$W^*$-algebra purely in terms of commutants: $\A$ is $W^*$
iff $\A=\A''$. 

This allows to easily  generalize the notion of `algebra generated 
by observables': Let $\{O_{\lambda}\}$ be a set of self-adjoint 
elements in $B(\H)$, then $\O:=\{O_{\lambda}\}''$ is called the 
($W^*$-) algebra generated by this set. This definition is natural 
since $\{O_{\lambda}\}''$ is easily seen to be the smallest 
$W^*$-algebra containing $\{O_{\lambda}\}$, for if 
$\{O_{\lambda}\}\subseteq\B\subseteq\O$ for some $W^*$-algebra $\B$, 
then taking the commutant twice yields 
%% BEGIN FOOTNOTE
$\B=\O$.\footnote{Note: for any $M\subseteq B(\H)$ definition 
(\ref{commutant}) immediately yields $M\subseteq M''$ and hence
$M'\supseteq M'''$ (by (\ref{commutant-inclusion})). But also 
$M'\subseteq M'''$ (by replacing $M\rightarrow M'$); therefore
$M'=M'''$ for any $M\subseteq B(\H)$.} 
%% END FOOTNOTE

Now we see that Dirac's requirement in the form (\ref{Dirac1})
directly translates to the general case if all algebras involved
(i.e. $\A$ and $\O$) are understood as $W^*$-algebras. Now we also 
know what a `complete set of (bounded) commuting observables' is, 
namely a set $\{A_{\lambda}\}\subseteq B(\H)$ whose generated 
$W^*$-algebra $\A:=\{A_{\lambda}\}''$ is maximal abelian: $\A=\A'$. 
This latter condition is again equivalent to the existence of a 
cyclic vector $\vert g\rangle\in\H$ for $\A$, where in infinite 
dimensions the definition of cyclic is that $\{\A\vert g\rangle\}$ 
is {\it dense} in (rather than equal to) $\H$.
It is also still true that there is an observable $A$ such that all
$\A_{\lambda}$ are functions (in an appropriate sense, not just 
polynomials of course) of $A$~\cite{vonNeumann}. But since $A$'s 
spectrum may be (partially) 
continuous, there is no direct interpretation of a `simple' spectrum
as in finite dimensions. Rather, one now defines simplicity of the 
spectrum of $A$ by the existence of a cyclic vector for $\A=\{A\}''$.

Now we come to our final reformulation of Dirac's condition. 
Namely, looking at (\ref{Dirac1}), we may ask whether we could not 
reformulate the existence of such a maximal abelian $\A$ purely 
in terms of the algebra of observables $\O$ alone. This is indeed 
possible. 
We have $\A\subseteq\O\Rightarrow \O'\subseteq\A'=\A\subseteq\O$, 
hence $\O'\subseteq\O$. Since $\O=\O''$ the last condition is 
equivalent to saying that $\O'$ is abelian ($\O'\subseteq\O''$), or 
to saying that $\O'$ is the center $\O^c$ of $\O$, since by
(\ref{commutant}) and (\ref{center}) the center can be written as 
$\O^c=\O\cap\O'$. Now, conversely, it was shown in \cite{Jauch2}
that an abelian $\O'$ implies the existence of a maximal abelian 
$\A\subseteq\O$. Hence we have the following alternative 
formulation of Dirac's requirement, first spelled out, 
independently of (\ref{Dirac1}), by Wightman~\cite{Wightman}, 
who called it the `hypothesis of commutative superselection rules':
\begin{equation}
\fbox{\hbox{Dirac's requirement, $2^{\rm nd}$ version:
$\O'$ is abelian}}
\label{Dirac2}
\end{equation}

There are several interesting ways to interpret this condition.
From its derivation we know that it is equivalent to the existence 
of a maximal abelian $\A\subseteq\O$. But we can in fact make  
an apparently stronger statement, which also relates to the 
earlier footnote~\ref{maximality}, namely: (\ref{Dirac2})
is equivalent to the condition, that {\it any} abelian 
$\A\subseteq\O$ that is maximal in $\O$, i.e. satisfies $\A=\A'\cap\O$, 
is also maximal in 
%% BEGIN FOOTNOTE
$B(\H)$.\footnote{Proof: We need to show that 
$(\hbox{$\O'$ abelian})\Leftrightarrow (\A=\A'\cap\O\Rightarrow
\A=\A')$.
`$\Rightarrow$': $\O'$ abelian implies $\O'\subseteq \O''=\O$
and $\A\subseteq\O$ implies $\O'\subseteq\A'$, so that  
$\O'\subseteq \A'\cap\O$. Hence $\A=\A'\cap\O$ implies 
$\O'\subseteq \A$, which implies $\A'\subseteq\O''=\O$, and hence
$\A=\A'$. `$\Leftarrow$': $(\A=\A'\cap\O\Rightarrow \A=\A')$
is equivalent to $\A'\subseteq\O$, which implies
$\O'\subseteq\A''=\A$ and hence that $\O'$ is abelian.}
%% END FOOTNOTE

\subsection{Dirac's condition and gauge symmetries}
Another way to understand (\ref{Dirac2}) is via its limitations on 
{\it gauge-symmetries}. To see this, we mention that any $W^*$-algebra 
is generated by its unitary elements. Hence $\O'$ is 
generated by a set $\{U_{\lambda}\}$ of unitary operators. 
Each $U_{\lambda}$ commutes with {\it all} observables and 
therefore generates a one-parameter group of gauge-transformations. 
Condition (\ref{Dirac2}) is then equivalent to saying that the 
total gauge group, which is generated by all $U_{\lambda}$, is 
{\it abelian}. Note also that an abelian $\O'$ implies that the 
gauge-algebra, $\{U_{\lambda}\}''=\O'$, is contained in the 
observables, $\O'\subseteq\O''=\O$, so that $\O'=\O^c$.
From this one can infer the following central statement:
\begin{equation}
\hbox{\fbox{\parbox{7.2truecm}{\sloppy 
Dirac's requirement implies that gauge- and 
sectorial structures are fully determined by the 
center $\O^c$ of the algebra of observables $\O$.
%
%The gauge- and sectorial structures compatible with Dirac's 
%requirement are fully determined by the center $\O^c$ of the 
%algebra of observables $\O$.
}}}
\label{Dirac3}
\end{equation}

To see in what sense this is true we remark that for $W^*$-algebras 
we can simultaneously diagonalize all observables in $\O^c$. That 
means that we can write $\H$ in an essentially unique way as direct 
integral over the real line of Hilbert spaces $\H(\lambda)$ using 
some (Lebesgue-Stieltjes-) measure $\sigma$:
\begin{equation}
\H=\int_{\reals}^{\oplus}d\sigma(\lambda)\,\H(\lambda).
\label{ssrintegral}
\end{equation}
Operators in $\O$ respect this decomposition in the sense that 
each $O\in\O$ acts on $\H$ componentwise via some bounded operator 
$O(\lambda)$ on $\H(\lambda)$. If $O\in\O^c$ then each $O(\lambda)$ 
is a multiple $\phi(\lambda)\in\complex$ of the unit operator. Moreover,   
the set of all $\{\O(\lambda)\}$ induced from $\O$ for each fixed 
$\lambda$ acts irreducibly on 
%% BEGIN FOOTNOTE
$\H(\lambda)$.\footnote{It is this 
irreducibility statement which depends crucially on the fulfillment 
of Dirac's requirement. In general, the $O(\lambda)$'s will act 
irreducibly on $\H(\lambda)$ for each $\lambda$, iff 
$\O^c$ is maximal abelian {\it in} $\O'$, i.e., iff 
$\O^c=(\O^c)'\cap\O'$. But we already saw that (\ref{Dirac2}) also 
implies $\O^c=\O'$ so that this is fulfilled.} 
%% END FOOTNOTE
Hence, provided that Dirac's requirement is satisfied, 
(\ref{ssrintegral}) is the generally valid version of (\ref{ssrsum}). 
The notion of disjointness now acquires an intuitive meaning: 
two states $\vert\Psi_1\rangle$ and $\vert\Psi_2\rangle$ are 
separated by a superselection rule (are disjoint), iff their 
component-state-functions 
$\lambda\rightarrow\vert\psi_1(\lambda)\rangle$ and 
$\lambda\rightarrow\vert\psi_2(\lambda)\rangle$ have 
disjoint support on $\reals$ (up to measure-zero sets). 
Note that by spectral decomposition 
the superselection observables can be decomposed into the projectors
in $\O^c$, which for (\ref{ssrintegral}) are all given by 
multiplications with characteristic functions $\chi(\lambda)$ for 
$\sigma$-measurable sets in $\reals$.

\subsubsection{Non-abelian gauge groups}
We have seen that the fulfillment of Dirac's requirement 
allows to give a full structural characterisation for the 
spaces of (pure) states and observables. How general is this 
result? Does it exclude cases of physical interest? At first 
glance this seems indeed to be the case: just consider a 
situations  with non-abelian gauge groups; for example, the 
quantum mechanical system of $n>2$ identical spinless 
particles with $n$-particle Hilbert space $\H=L^2(\reals^{3n})$ 
on which the permutation group $G=S_n$ of $n$ objects acts in 
the obvious way by unitary operators $U(g)$. That these particles 
are identical means that observables must commute with each $U(g)$. 
Without further restrictions on observables one would thus define 
$\O:=\{U(g),\,g\in G\}'$. Hence $\O'$ is the $W^*$-algebra 
generated by all $U(g)$, which is clearly non-abelian, thus 
violating~(\ref{Dirac2}). But does this generally imply that general 
particle statistics cannot be described in a quantum-mechanical setting 
which fulfills Dirac's requirement? The answer to this question 
is `no'. Let us explain why.

If we decompose $\H$ according to the unitary, irreducible 
representations of $G$ we obtain (\cite{GMN}\cite{Giulini1}) 
\begin{equation}
\H=\bigoplus_{i=1}^{p(n)}\H_i\,,
\label{stat1}
\end{equation}
where $i$ labels the $p(n)$ inequivalent, unitary, irreducible 
representations $D_i$ of $G$ of dimension $d_i$. Each $\H_i$ 
has the structure $\H_i\cong\complex^{d_i}\otimes{\t\H}_i$, where $G$ 
acts irreducibly via $D_i$ on $\complex^{d_i}$ and trivially on ${\t\H}_i$
whereas $\O$ acts irreducibly via some $*$-representation $\pi_i$ 
on ${\t\H}_i$ and trivially on $\complex^{d_i}$. $\pi_i$ and $\pi_j$ are 
inequivalent if $i\not =j$. Hence we see that $\H_i$ furnishes an 
irreducible representation for $\O$, iff $d_i=1$, i.e., for the 
Bose and Fermi sectors only. Pure states from these sectors are 
just the rays in the corresponding $\H_i$. In contrast, for $d_i>1$, 
given a non-zero vector $\vert\phi\rangle\in{\t\H}_i$, all non-zero 
vectors in the $d_i$-dimensional subspace
$\complex^{d_i}\otimes\vert\phi\rangle\subset\H_i$ define the {\it same} 
pure state, i.e., the same expectation-value-functional on $\O$.
Furthermore, a vector in $\H_i\cong\complex^{d_i}\otimes{\t\H}_i$ which is  
not a pure tensor product defines a non-pure state, since the 
restriction of $O\in\O$ to $\H_i$ is of the form ${\bf 1}\otimes\t O$,
which means that a vector in $\H_i$ defines a state given by the 
reduced density matrix obtained by tracing over the left 
(i.e. $\complex^{d_i}$) state space. From elementary quantum
mechanics we know 
that the resulting state is pure, iff the vector in $\H_i$ was a pure 
tensor product (i.e. of rank one).
Hence in those $\H_i$ where $d_i>1$ not all vectors correspond to pure 
states, and those which do represent pure states in a redundant 
fashion by higher dimensional subspaces, sometimes called `generalized 
rays' in the older literature on parastatistics~\cite{Messiah}.

However, the factors $\complex^{d_i}$ are completely redundant as far as 
physical information is concerned, which is already fully encoded in
the irreducible representations $\pi_i$ of $\O$ on ${\t\H}_i$;
no further physical information is contained in $d_i$-fold
repetitions of $\pi_i$. Hence we can define a new, truncated 
Hilbert space 
\begin{equation}
\t\H:=\bigoplus_{i=1}^{p(n)}{\t\H}_i\,.
\label{stat2}
\end{equation}
This procedure has also been called `elimination of the generalized 
ray' in the older literature on parastatistics~\cite{HT} -- see also 
\cite{Giulini1} for a more recent discussion of this point. 
Since every pure state in $\H$ is also contained in $\t\H$, just 
without repetition, these two sets are called `phenomenological 
equivalent' in the literature on QFT~(e.g. in chapter 6.1.C of 
\cite{BLOT}).
The point is that pure states are now faithfully labelled by rays in 
the ${\t\H}_i$ and that $\O'$ -- where the commutant is now taken in 
$B(\t\H)$ rather than $B(\H)$ -- is generated by ${\bf 1}$ and the 
$p(n)$ (commuting!) projectors into the ${\t\H}_i$'s. Hence Dirac's 
requirement is satisfied. But clearly the original gauge group has 
no action on $\t\H$ anymore, but there is also no physical reason 
why one should keep 
%% BEGIN FOOTNOTE
it.\footnote{In \cite{GMN} Dirac's requirement 
together with the requirement that the physical Hilbert space must 
carry an action of the gauge group has been used to ``prove'' the 
absence of parastatistics. In our opinion there seems to be no physical  
reason to accept the second requirement and hence the ``proof''; 
compare~\cite{HT} and~\cite{Giulini1}.}
%% END FOOTNOTE
It served to define $\O$, but then only its irreducible
representations $\pi_i$ are of interest. Only a residual 
action of the center of $G$ still exists, but the gauge 
group generated by the projectors into the ${\t\H}_i$ 
consists in fact of the continuous group of $p(n)$ copies 
of $U(1)$, one global phase change for each sector.
Its meaning is simply to induce the separation into the 
different sectors $({\t\H}_i,\pi_i)$, and that in 
accordance with Dirac's requirement. 

To sum up, we have seen that even if a theory is initially 
formulated via non-abelian gauge groups, we can give it 
a physically equivalent formulation that has at most a 
residual abelian gauge group left and hence obeys Dirac's 
requirement. Hence the `obvious' counterexamples to Dirac's 
requirement turn out to be harmless. This is generally true 
in quantum mechanics, but in quantum field theory there are 
genuine possibilities to violate Dirac's condition which
we will ignore 
%% BEGIN FOOTNOTE
here.\footnote{An abelian $\O'$ implies that $\O$ 
is a von Neumann algebra of type~I (see \cite{Dixmier}, chapter~8) 
whereas truly infinite systems in QFT are often described by 
type~III algebras.}
%% END FOOTNOTE

\section{Superselection Rules via Symmetry Requirements}
The requirement that a certain group must act on the set of all
physical states is often the (kinematical) source of superselection
rules. Here I wish to explain the structure of this argument.

Note first that in quantum mechanics we identify the states of a 
closed system with rays and not with vectors which represent them 
(in a redundant fashion).
It is therefore not necessary to require that a symmetry group $G$ 
acts on the Hilbert space $\H$, but rather it is sufficient that 
it acts on $\PH$, the space of rays, via so-called ray-representations.
Mathematically this is a non-trivial relaxation since not every 
ray-representation of a symmetry group $G$ (i.e. preserving the ray products) 
lifts to a unitary action of $G$ on $\H$. What may go wrong is not that 
for a given $g\in G$ we cannot find a unitary (or anti-unitary) 
operator ${\op U}_g$ on $\H$; that is assured by Wigner's 
theorem (see \cite{Bargmann} for a proof).  
Rather, what may fail to be possible is that we can choose the 
${\op U}_g$'s in such a way that we have an {\it action}, i.e., that 
${\op U_{g_1}}{\op U_{g_2}}=U_{g_1g_2}$. As is well known, this is 
precisely what happens for the implementation of the Galilei group in 
ordinary quantum mechanics. Without the admission of ray 
representations we would not be able to say that ordinary quantum 
mechanics is Galilei invariant.   

To be more precise, to have a ray-representation means that for each 
$g\in G$ there is a 
%% BEGIN FOOTNOTE
unitary\footnote{For simplicity we ignore 
anti-unitary transformations. They cannot arise if, for example, 
$G$ is connected.} 
%% END FOOTNOTE
transformation ${\op U}_g$ which, instead 
of the usual representation property, are only required
to satisfy the weaker condition
\begin{equation}
{\op U}_{g_1}{\op U}_{g_2}=\exp(i\xi(g_1,g_2))\,{\op U}_{g_1g_2},
\label{ray}
\end{equation}
for some function $\xi:G\times G\rightarrow\reals$, called multiplier
exponent, 
%% BEGIN FOOTNOTE
satisfying\footnote{\label{eqzero}The following conditions might seem 
a little too strong, since it would be sufficient to require 
the equalities in (20) and (\ref{cocycle}) only mod 
$2\pi$; this also applies to (\ref{redef}). But for our application in 
section~4 it is more convenient to work with strict equalities, 
which in fact implies no loss of generality; 
compare \cite{Raghunathan}.}
%% END FOOTNOTE
\begin{eqnarray}
\xi(1,g)=\xi(g,1)&=&0, \\
\label{zero}
\xi(g_1,g_2)-\xi(g_1,g_2g_3)+\xi(g_1g_2,g_3)-\xi(g_2,g_3)&=&0.
\label{cocycle}
\end{eqnarray}
The second of these conditions is a direct consequence of 
associativity: ${\op U_{g_1}}({\op U_{g_2}}{\op U_{g_3}})=
({\op U_{g_1}}{\op U_{g_2}}){\op U_{g_3}}$.
Obviously these maps project to an action of $G$ on $\PH$. Any other 
lift of this action on $\PH$ onto $\H$ is given by a redefinition
${\op U}_g\rightarrow {\op U}'_g:=\exp(i\gamma(g)){\op U}_g$, for some 
function $\gamma:G\rightarrow\reals$ with $\gamma(1)=0$, 
resulting in new multiplier exponents
\begin{equation}
\xi'(g_1,g_2)=\xi(g_1,g_2) +\gamma(g_1)-\gamma(g_1g_2)+\gamma(g_2),
\label{redef}
\end{equation}
which again satisfy (20) and (\ref{cocycle}). The ray 
representations ${\op U}$ and
${\op U'}$ are then said to be equivalent, since the projected actions on
$\PH$ are the same. We shall also say that two multiplier exponents
$\xi,\xi'$ are equivalent if they satisfy (\ref{redef}) for some $\gamma$.

We shall now see how the existence of inequivalent multiplier exponents, 
together with the requirement that the group should 
act on the space of physical states, may clash with the 
superposition principle and thus give rise to superselection rules.  
For this we start from two Hilbert spaces $\H'$ and $\H''$ and 
actions of a symmetry group $G$ on $\PH'$ and $\PH''$, i.e., ray 
representations ${\op U'}$ and ${\op U''}$ on $\H'$ and $\H''$ up to 
equivalences (\ref{redef}). We consider $\H=\H'\oplus\H''$ and ask under 
what conditions does there exist an action of $G$ on $\PH$ which 
restricts to the given actions on the subsets $\PH'$ and $\PH''$. 
Equivalently: when is ${\op U}={\op U'}\oplus{\op U''}$ a ray 
representation of $G$ on $\H$ for some choice of ray-representations 
${\op U}'$ and ${\op U}''$ within their equivalence class? To answer 
this question, we consider 
\begin{eqnarray}
{\op U}_{g_1}{\op U}_{g_2}
&=& ({\op U}'_{g_1}\oplus{\op U}''_{g_1})
    ({\op U}'_{g_2}\oplus{\op U}''_{g_2}) \nonumber\\
&=& \exp(i\xi'(g_1,g_2)){\op U}'_{g_1g_2}
    \oplus\exp(\xi''(g_1,g_2)){\op U}''_{g_1g_2}
\label{sum}
\end{eqnarray}
and note that this can be written in the form (\ref{ray}), for some choice 
of $\xi',\xi''$ within their equivalence class, iff the phase factors can 
be made to coincide, that is, iff $\xi'$ and $\xi''$ are equivalent.
This shows that there exists a ray-representation on
$\H$ which restricts to the given equivalence classes of given ray
representations on $\H'$ and $\H'$, iff the multiplier exponents
of the latter are equivalent. Hence, if the multiplier 
exponents $\xi'$ and $\xi''$ are {\it not} equivalent, the action of $G$
cannot be extended beyond the disjoint union $\PH'\cup\PH''$.
Conversely, {\it if} we require that the space of physical states 
must support an action of $G$, then non-trivial superpositions 
of states in $\H'$ and $\H''$ must be excluded from the space of 
(pure) physical states. 

This argument shows that if we insist of implementing $G$ 
as symmetry group, superselection rules are sometimes 
unavoidable. A formal trick to avoid them would be not to 
require $G$, but a slightly larger group, $\b G$, to act on the 
space of physical states. $\b G$ is chosen to be the group 
whose elements we label by $(\theta,g)$, where $\theta\in\reals$, 
and the multiplication law is 
\begin{equation}
{\b g}_1{\b g}_2 = (\theta_1,g_1)(\theta_2,g_2)
                 = (\theta_1+\theta_2+\xi(g_1,g_2),g_1g_2).
\label{extension}
\end{equation}
It is easy to check that the elements of the form $(\theta, 1)$ 
lie in the center of $\b G$ and form a normal subgroup 
$\cong\reals$ which we call $Z$. Hence $\b G/Z=G$ but $G$ 
need not be a subgroup of $\b G$. $\b G$ is a central $\reals$ 
%% BEGIN FOOTNOTE
extension\footnote{Had we defined the multiplier exponents mod 
$2\pi$ (compare footnote~\ref{eqzero}) then we would have obtained a
$U(1)$ 
extension, which would suffice so far. But in the next section 
we will definitively need the $\reals$ extension as symmetry group  
of the extended classical model discussed there.}
%% END FOOTNOTE
of $G$~ (see e.g. \cite{Raghunathan}). 
Now a ray-representation ${\op U}$ 
of $G$ on $\H$ defines a proper representation $U$ of
$\b G$ on $\H$ by setting 
\begin{equation}
U_{(\theta,g)}:=\exp(i\theta){\op U_g}.
\label{newrep}
\end{equation}
Then $\b G$ is properly represented on $\H'$ and $\H''$ and hence 
also on $\H=\H'\oplus\H''$. The above phenomenon is mirrored here 
by the fact that $Z$ acts trivially on $\PH'$ and $\PH''$
but non-trivially on $\PH$, and the superselection structure
comes about by requiring physical states to be fixed points of 
$Z$'s action.

\section{Bargmann's Superselection Rule}
An often mentioned textbook example where a particular 
implementation of a symmetry group allegedly clashes with the 
superposition principle, such that a superselection rule results, 
is Galilei invariant quantum mechanics (e.g. \cite{GP}; see
also Wightman's review \cite{Wightman2}). We will discuss this 
example in detail for the general multi-particle case. (Textbook 
discussions usually restrict to one particle, which, due to Galilei
invariance, must necessarily be free.) It will serve as a test case to 
illustrate the argument of the previous chapter and also to formulate 
our critique. Its physical significance is limited by the fact that the 
particular feature of the Galilei group that is responsible for the 
existence of the mass superselection rule ceases to exist if we replace 
the Galilei group by the Poincar\'e group (i.e. it is unstable under 
`deformations'). But this is not important for our 
%% BEGIN FOOTNOTE
argument.
\footnote{In General Relativity, where the total mass can be 
expressed as a surface integral at `infinity', the issue of mass 
superselection comes up again; see e.g. \cite{GKZ} and
\cite{Dominguez}.}
%% END FOOTNOTE
Let now $G$ be the Galilei group, an element of which is parameterized
by $(R,\vec v,\vec a,b)$, with $R$ a rotation matrix in $SO(3)$,
$\vec v$ the boost velocity, $\vec a$ the spatial translation,
and $b$ the time translation. Its laws of multiplication and inversion 
are respectively given by 
\begin{eqnarray}
g_1g_2 &=& (R_1,\vec v_1,\vec a_1,b_1)(R_2,\vec v_2,\vec a_2,b_2)\nonumber\\
       &=& (R_1R_2\,,\,\vec v_1+R_1\cdot\vec v_2\,,\,a_1+R_1\cdot\vec a_2+
            \vec v_1b_2\,,\,b_1+b_2),
\label{mult}\\
g^{-1}&=& (R,\vec v,\vec a,b)^{-1}=(R^{-1},\,-R^{-1}\cdot\vec v\,,\, 
-R^{-1}\cdot(\vec a-\vec vb)\,,\,-b).
\label{inv}
\end{eqnarray} 
We consider the Schr\"odinger
equation for a system of $n$ particles of positions $\vec x_i$,
masses $m_i$, mutual distances $r_{ij}:=\Vert\vec x_i-\vec x_j\Vert$
which interact via a Galilei-invariant potential $V(\{r_{ij}\})$,
so that the Hamilton operator becomes
${\op H}=-\hbar^2\sum_i\frac{\Delta_i}{2m_i}+V$. The Hilbert space
is $\H=L^2(\reals^{3n},d^3\vec x_1\cdots d^3\vec x_n)$.

$G$ acts on the space 
$\{\hbox{configurations}\}\times\{\hbox{times}\}\cong\reals^{3n+1}$ 
as follows: Let $g=(R,\vec v,\vec a,b)$, then 
$g(\{\vec x_i\},t):=(\{R\cdot\vec x_i+\vec vt+\vec a\}\,,\,t+b)$.
Hence $G$ has the obvious left action on complex-valued functions on 
$\reals^{3n+1}$: $(g,\psi)\rightarrow\psi\circ g^{-1}$. However, these
transformations do {\it not} map solutions of the Schr\"odinger equations
into solutions. But, as is well known, this can be achieved by
introducing an $\reals^{3n+1}$-dependent phase factor (see e.g. 
\cite{Giulini2} for a general derivation). 
We set $M=\sum_i m_i$ for the total mass and 
$\vec r_c=\frac{1}{M}\sum_im_i\vec x_i$ for the center-of-mass. 
Then the modified transformation, ${\op T}_g$, which maps solutions 
(i.e. curves in $\H$) to solutions, is given by
\begin{equation}
{\op T}_g\psi(\{\vec x_i\},t):=
\exp\left(\ihbar M[\vec v\cdot(\vec r_c-\vec a)
-\shalf\vec v^2(t-b)]\right)\,\psi(g^{-1}(\{\vec x_i\}, t)).
\label{trans}
\end{equation}
However, due to the modification, these transformations have lost the 
property to define an action of $G$, that is, we do {\it not} have 
${\op T}_{g_1}\circ {\op T}_{g_2}={\op T}_{g_1g_2}$. Rather, a straightforward 
calculation using (\ref{mult}) and (\ref{inv}) leads to 
\begin{equation}
{\op T}_{g_1}\circ{\op T_{g_2}}
=\exp(i\xi(g_1,g_2))\,{\op T}_{g_1g_2},
\label{raytrans}
\end{equation}
with non-trivial multiplier exponent
\begin{equation}
\xi(g_1,g_2)=
\Mhbar({\vec v}_1\cdot R_1\cdot{\vec a}_2+\shalf{\vec v}_1^2b_2).
\label{Gphase}
\end{equation} 
Although each ${\op T}_g$ is a mapping of {\it curves} in 
$\H$, it also defines a unitary transformation on $\H$ itself. 
This is so because the equations of motion define a bijection 
between solution curves and initial conditions at, say, $t=0$,
which allows to translate the map ${\op T}_g$ into a unitary 
map on $\H$, which we call ${\op U}_g$. It is given by  
\begin{equation}
{\op U}_g\psi(\{\vec x_i\})=
\exp\left(\ihbar M[\vec v\cdot(\vec r_c-\vec a)
+\shalf\vec v^2b]\right)\,
\exp(\ihbar{\op H}b)\psi(\{R^{-1}(\vec x_i-\vec a+\vec vb)\}),
\label{Galileitrans}
\end{equation}
and furnishes a ray-representation whose multiplier exponents
are given by (\ref{Gphase}). It is easy to see that the multiplier 
exponents are non-trivial, i.e., not removable by a redefinition 
(\ref{redef}). The quickest way to see this is as follows: suppose 
to the contrary that they were trivial and that hence (\ref{redef}) 
holds with $\xi'\equiv 0$. Trivially, this equation will continue to 
hold after restriction to any subgroup $G_0\subset G$. We choose for 
$G_0$ the abelian subgroup generated by boosts and space translations,
so that the combination $\gamma(g_1)-\gamma(g_1g_2)+\gamma(g_2)$
becomes symmetric in $g_1,g_2\in G_0$. But the exponent (\ref{Gphase})
stays obviously asymmetric after restriction to $G_0$. Hence no 
cancellation can take place, which contradicts our initial assumption.

The same trick immediately shows that the multiplier exponents are
inequivalent for different total masses $M$. Hence, by the general 
argument given in the previous chapter, if $\H'$ and $\H''$
correspond to Hilbert spaces of states with different overall masses 
$M'$ and $M''$, then the requirement that the Galilei group should act 
on the set of physical states excludes superpositions of states of 
different overall mass. This is Bargmann's superselection rule.

I criticize these arguments for the following reason: The dynamical 
framework that we consider here treats `mass' as parameter(s) which 
serves to specify the system. States for different overall masses are 
states of {\it different} dynamical systems, to which the superposition 
principle does not even potentially apply. In order to investigate a 
possible violation of the superposition principle, we must find a 
dynamical framework in which states of different overall mass are states 
of the {\it same} system; in other words, where mass is a dynamical  
variable. But if we enlarge our system to one where 
mass is dynamical, it is not at all obvious that the Galilei group 
will survive as symmetry group. We will now see that in fact it 
does not, at least for the simple dynamical extension which we
now discuss.

The most simple extension of the classical model is to maintain the 
Hamiltonian, but now regarded as function on an extended,  
$6n+2n$ - dimensional phase space with extra `momenta' $m_i$ and
conjugate generalized `positions' $\lambda_i$. Since the 
$\lambda_i$'s do not appear in the Hamiltonian, the $m_i$'s are
constants of motion. Hence the equations of motion for the 
$\vec x_i$'s and their conjugate momenta $\vec p_i$ are unchanged 
(upon inserting the integration constants $m_i$) and those of the 
new positions $\lambda_i$ are 
\begin{equation}
\dot\lambda_i(t)=\frac{\partial V}{\partial m_i}-\frac{\vec p_i^2}{2m_i^2},
\label{zetadot}
\end{equation}
which, upon inserting the solutions $\{\vec x_i(t),\vec p_i(t)\}$,
are solved by quadrature. 

Now, the point is that the new Hamiltonian equations of motion do not 
allow the Galilei group as symmetries anymore. But they do allow the 
$\reals$-extension $\b G$ as symmetries \cite{Giulini2}. Its 
multiplication law is given by (\ref{extension}), with $\xi$  as in 
(\ref{Gphase}). The action of $\b G$ on the extended space of
$\{\hbox{configurations}\}\times\{\hbox{times}\}$ is now given by 
\begin{eqnarray}
&&\b g(\{\vec x_i\},\{\lambda_i\},t)=
(\theta,R,\vec v,\vec a,b)(\{\vec x_i\},\{\lambda_i\},t)\nonumber\\
&&=(\{R\vec x_i+\vec v t+\vec a\}\,,\,
\{\lambda_i-(\hbarM\theta+\vec v\cdot R\cdot\vec x_i+\shalf\vec v^2t)\}
\,,\, t+b).
\label{newaction}
\end{eqnarray}
With (\ref{extension}) and (\ref{Gphase}) it is easy to verify that 
this defines indeed an action. Hence it also defines an action on 
curves in the new Hilbert space $\b\H:=L^2(R^{4n},d^{3n}\vec x d^n\lambda)$, 
given by 
\begin{equation}
{\b{\op T}}_{\b g}\psi:=\psi\circ {\b g}^{-1}\,,
\label{newtrans}
\end{equation}
which already maps solutions of the new Schr\"odinger equation to
solutions, {\it without} invoking non-trivial phase factors. 
This is seen as follows: 
Let $\Psi(\{{\vec x}_i\},\{\lambda_i\},t)\in\b\H$ and 
$\Phi(\{{\vec x}_i\},\{m_i\},t)$ its Fourier transform in the 
$(\lambda_i,m_i)$ arguments: 
\begin{equation}
\Phi(\{{\vec x}_i\},\{\lambda_i\},t)=(2\pi\hbar)^{-n/2}
\int_{\reals^n}d^nm\,\exp\left[\frac{i}{\hbar}\sum_{i=1}^n
m_i\lambda_i\right]\ \Phi(\{{\vec x}_i\},\{m_i\},t).
\label{Fourier}
\end{equation}
For each set of masses $\{m_i\}$ the function 
$\Phi_{\{m_i\}}(\{{\vec x}_i\},t):=\Phi(\{{\vec x}_i\},\{m_i\},t)$ 
satisfies the original Schr\"odinger equation. Since (\ref{newtrans}) 
does not mix different sets of $\{m_i\}$ it induces a map   
${\b{\op T}}^{\{m_i\}}_{\b g}$ for each such set:
\begin{eqnarray}
 {\b{\op T}}^{\{m_i\}}_{\b g}\Phi_{\{m_i\}}(\{{\vec x}_i\},t)
:& = &
 \exp\left[i\theta+\ihbar M\left(\vec v\cdot(\vec r_c-\vec a)-
 \shalf{\vec v}^2 (t-b)\right)\right]\
\nonumber\\
 &\times &
\Phi_{\{m_i\}}(g^{-1}(\{{\vec x}_i\},t))
\label{fourtrans}
\end{eqnarray}
Via the Fourier transform (\ref{Fourier}) we represent $\b\H$ as 
direct integral of ${\H}_{\{m_i\}}$'s, each of which isomorphic
to our old $\H=L^2(\reals^{3n},\,d^3{\vec x}_1\cdots d^3{\vec x}_n)$,
and on each of which (\ref{fourtrans}) defines a  unitary 
representation $U$ of $\b G$ the form (\ref{newrep}) with 
${\op U}_g$ the ray-representation (\ref{Galileitrans}). 
This shows how the much simpler transformation law (\ref{newtrans})
contains the more complicated one (\ref{trans}) upon writing $\b\H$ 
as a direct integral of vector spaces $\H_{\{m_i\}}$.

In the new framework the overall mass, $M$, is a dynamical variable, 
and it would make sense to state a superselection rule with  respect 
to it. But now $\b G$ rather than $G$ is the dynamical symmetry group, 
which acts by a proper unitary representation on $\b\H$, so that the  
requirement that the dynamical symmetry group should act on the space 
of physical states will now not lead to any superselection rule. Rather, 
the new and more physical interpretation of a possible superselection 
rule for $M$ would be that we cannot localize the system in the 
coordinate conjugate to overall mass, which we call 
$\Lambda$, i.e., that only the {\it relative} new positions 
$\lambda_i-\lambda_j$ are 
%% BEGIN FOOTNOTE
observable.\footnote{A  system 
$\{(\t\lambda_i,\t m_i\})$ of canonical coordinates including 
$M=\sum_im_i$ is e.g. $\t\lambda_1:=\lambda_1$, ${\t m}_1=M$ and 
$\t\lambda_i=\lambda_i-\lambda_1$, ${\t m}_i=m_i$ for $i=2...n$. Then 
$\Lambda=\t\lambda_1$.}
%% END FOOTNOTE
(This is so because $M$ generates translations of equal amount in all 
$\lambda_i$.) But this would now be a contingent physical property rather 
than a mathematical necessity. Note also that in our 
dynamical setup it is inconsistent to just state that $M$ generates 
gauge symmetries, i.e. that $\Lambda$ corresponds to a physically non 
existent degree of freedom. For example, a motion in real time
along $\Lambda$ requires a non-vanishing action (for non-vanishing 
$M$), due to the term $\int dt\, M\dot\Lambda$ in the expression for 
the action. 

If decoherence were to explain the (ficticious) mass superselection
rule, it  would be due to a dynamical instability (as explained
in \cite{Joos}) of those states which are more or less localized in
$\Lambda$. Mathematically this effect would be modelled by
effectively removing the projectors onto  $\Lambda$-subintervalls 
from the algebra of observables, thereby putting $M$ (i.e. its 
projectors) into the center of $\O$. Such a non-trivial center 
should therefore be thought of as resulting from an 
approximation-dependent idealisation.
\vfill\eject

\section{Charge Superselection Rule}
In the previous case I said that superselection rules should be
stated within a dynamical framework including  as dynamical
degree of freedom the direction generated by the superselected 
quantity. What is this degree of freedom in the case of a
superselected electric charge and how does it naturally appear
within the dynamical setup? What is its relation to the 
Coulomb field whose r\^ole in charge-decoherence has been suggested 
in \cite{GKZ}? In the following discussion I wish to investigate 
into these questions by looking at the Hamiltonian formulation of 
Maxwell's equation and the associated canonical quantization. 

In Minkowski space, with preferred coordinates $\{x^{\mu}=(t,x,y,z)\}$ 
(laboratory rest frame), we consider the spatially finite region 
$Z=\{(t,x,y,z):\, x^2+y^2+z^2\leq R^2\}$. $\Sigma$ denotes the intersection 
of $Z$ with a slice $t=\hbox{const.}$ and $\partial\Sigma=:S_R$ its 
boundary (the laboratory walls). Suppose we wish to solve Maxwell's equations 
within $Z$, allowing for charged solutions. It is well known that in order 
for charged configurations to be stationary points of the action, the 
standard action functional has to be supplemented by certain surface 
terms (see e.g. \cite{Gervais-Zwanziger}) which involve new fields
on the boundary, which we call $\lambda$ and $f$, and which represent 
a pair of canonically conjugate variables in the Hamiltonian sense.
On the laboratory walls, $\partial\Sigma$, we put the boundary 
conditions that the normal component of the current and the
tangential components of the magnetic field vanish. Then the appropriate 
boundary term for the action reads
\begin{equation}
\int_Z dt\, d\omega (\dot\lambda+\phi)f,
\label{boundary}
\end{equation}
where $\phi$ is the scalar potential and $d\omega$ the measure on the 
spatial boundary 2-sphere rescaled to unit radius.
Adding this to the standard action functional and expressing all 
fields on the spatial boundary by their multipole moments (so that 
integrals $\int_{\partial\Sigma} d\omega\,R^2$, $d\omega=$ measure on   
unit sphere, become $\sum_{lm}$), one arrives at a Hamiltonian function
\begin{equation}
H=\int_{\Sigma}\left[\shalf(\vec E^2+(\vec\nabla\times\vec A)^2)+
\phi(\rho-\vec\nabla\cdot\vec E)-\vec A\cdot\vec j\right]
+\sum_{lm}\phi_{lm}(E_{lm}-f_{lm}).
\label{Hamiltonian}
\end{equation}
Here the pairs of canonically conjugate  variables are 
$(\vec A(\vec x),-\vec E(\vec x))$ and $(\lambda_{lm},f_{lm})$, 
and $E_{lm}$ are the multipole components of
$\vec n\cdot\vec E$, 
\begin{equation}
E_{lm}:=\int_{\partial\Sigma}d\omega R^2\, Y_{lm}{\vec n\cdot \vec E},
\label{multipole}
\end{equation}
where $\vec n$ is the normal to $\partial\Sigma$.
The scalar potential $\phi$ has to be considered as Lagrange multiplier. 
With the given boundary conditions the Hamiltonian is differentiable 
with respect to all the canonical 
%% BEGIN FOOTNOTE
variables\footnote{This would not 
be true without the additional surface term (\ref{boundary}). 
Without it one does not simply obtain the wrong Hamiltonian 
equations of motions, but none at all! Concerning the Langrangean
formalism one should be aware that the Euler-Lagrange equations 
may formally admit solutions (e.g. with long-ranged (charged) 
fields) which are outside the class of functions which one used 
in the variational principle of the action (e.g. rapid fall-off).  
Such solutions are not stationary points of the action and their 
admittance is in conflict with the variational principle unless 
the expression for the action is modified by appropriate boundary 
terms.}
%% END FOOTNOTE
and leads to the following  equations of motion
\begin{eqnarray}
\dot{\vec A} 
&=& \frac{\delta H}{\delta (-\vec E)}
=-\vec E-\vec\nabla\phi\,,                               
\label{motion1}                                       \\
-\dot{\vec E} 
&=&-\frac{\delta H}{\delta\vec A}
=\vec j-\vec\nabla\times\vec (\vec\nabla\times\vec A)\,,
\label{motion2}                                       \\
{\dot\lambda}_{lm} 
&=& \frac{\partial H}{\partial f_{lm}}=-\phi_{lm}\,,     
\label{motion3}                                       \\
{\dot f}_{lm}
&=& -\frac{\partial H}{\partial\lambda_{lm}}=0\,.        
\label{motion4}                                       
\end{eqnarray}
These are supplemented by the equations which one obtains by varying 
with respect to the scalar potential $\phi$, which, as already said, 
is considered as Lagrange multiplier. Varying first with respect to
$\phi(\vec x)$ (i.e. within $\Sigma$) and then with respect to 
$\phi_{lm}$ (i.e. on the boundary $\partial\Sigma$), one obtains

\begin{eqnarray}
G(\vec x):&=&\vec\nabla\cdot\vec E(\vec x)-\rho(\vec x)=0,
\label{constraint1}                                      \\
G_{lm}:&=& E_{lm}-f_{lm}=0.
\label{constraint2}                                      
\end{eqnarray}
These equations are constraints (containing no time derivatives) which, 
once imposed on initial conditions, continue to hold due to the 
equations of motion.
%% BEGIN FOOTNOTE
\footnote{Equation (\ref{motion2}) together with 
charge conservation, $\dot\rho+\vec\nabla\cdot\vec j=0$, shows that 
(\ref{constraint1}) is preserved in time, and (\ref{motion4}) together 
with the boundary condition that $\vec n\cdot \vec j$ and 
$\vec n\times(\vec\nabla\times\vec A)$  vanish on $\partial\Sigma$
show that (\ref{constraint2}) is preserved in time.} 
%% END FOOTNOTE

This ends our discussion of the classical theory. The point 
was to show that it leaves no ambiguity as to what its dynamical degrees 
of freedom are, and that we had to include the variables $\lambda_{lm}$
along with their conjugate momenta $f_{lm}$ in order to gain consistency 
with the existence of charged configurations. The physical interpretation 
of the $\lambda_{lm}$'s  is not obvious. Equation (\ref{motion3}) merely 
relates their time derivative to the scalar potential's multipole moments 
on the boundary, which are clearly highly non-local quantities. The 
interpretation of the $f_{lm}$'s follow from (\ref{constraint2})
and the definition of $E_{lm}$, i.e. they are the multipole moments 
of the electric flux distribution 
$\varphi(\vec n):=R^2\vec n\cdot\vec E(R^2\vec n)$. 
In particular, for $l=0=m$ we have 
\begin{equation}
f_{00}=(4\pi)^{-\frac{1}{2}}\, Q,  
\label{charge}
\end{equation}
where $Q$ is the total charge of the system. Hence we see that the total 
charge  generates motions in $\lambda_{00}$. But this means that the 
degree of freedom labelled by $\lambda_{00}$ truly exists (in the
sense of the theory). For example, a motion along $\lambda_{00}$ will
cost a non-vanishing amount of action 
$\propto Q(\lambda_{00}^{\rm final}-\lambda_{00}^{\rm initial})$. 
A declaration that $\lambda_{00}$ really labels only a gauge degree
of freedom is {\it incompatible} with the inclusion of charged states.
Similar considerations apply of course to the other values of $l,m$. 
But note that this conclusion is independent of the radius $R$ of the 
spatial boundary 2-sphere $\partial\Sigma$. In particular, it 
continues to hold in the limit $R\rightarrow\infty$. We will not 
consistently get rid of physical degrees of freedom that way, even if we 
agree that realistic physical measurements will only detect field values 
in bounded regions of space-time. See \cite{Giulini3} for more 
discussion on this point and the distinction between proper 
symmetries and gauge symmetries. 

It should be obvious how these last remarks apply to the statement of a 
charge superselection rule. Without entering the technical issues 
(see e.g. \cite{Strocchi-Wightman}), its basic ingredient is Gauss' law
(for operator-valued quantities), locality of the electric field and 
causality. That $Q$ commutes with all (quasi-) local observables then 
follows simply from writing $Q$ as surface integral of the local flux 
operator $R^2\vec n\cdot\hat{\vec E}$, and the observation that the
surface may be taken to lie in the causal complement of any bounded 
space-time region. Causality then implies commutativity with any 
local observable.

In a heuristic Schr\"odinger picture formulation of QED one 
represents states $\Psi$ by functions of the configuration 
variables $\vec A(\vec x)$ and $\lambda_{lm}$. The momentum operators  
are obtained as usual: 
\begin{eqnarray}
-\vec E(\vec x)&\longrightarrow & 
-{\rm i}\frac{\delta}{\delta\vec A(\vec x)}\,, 
\label{quant1}\\
f_{lm}&\longrightarrow & -{\rm i}\frac{\partial}{\partial\lambda_{lm}}\,.
\label{quant2}
\end{eqnarray}
In particular, the constraint 
(\ref{constraint2}) implies the statement that on physical 
states $\Psi$ we 
%% BEGIN FOOTNOTE
have\footnote{Clearly all sorts of points are simply 
sketched over here. For example, charge quantization presumably means 
that $\lambda_{00}$ should be taken with a compact range, which in turn 
will modify (\ref{quant2}) and (\ref{Psi}). But this is irrelevant to 
the point stressed here.}
%% END FOOTNOTE
\begin{equation}
\hat Q\Psi=-{\rm i}\sqrt{4\pi}
\frac{\partial}{\partial\lambda_{00}}\Psi\,.
\label{Psi}
\end{equation}
This shows that a charge superselection rule is equivalent to the
statement that we cannot localize the system in its $\lambda_{00}$
degree of freedom. Removing {\it by hand} the multiplication
operator 
$\lambda_{00}$ (i.e. the  projectors onto $\lambda_{00}$-intervals) 
from our observables clearly makes $Q$ a central element in the 
remaining algebra of observables. But what is the physical justification 
for this removal? Certainly, it is valid FAPP if one restricts to local 
observations in space-time. To state that this is a {\it fundamental} 
restriction, and not only an approximate one, is equivalent to saying 
that for some fundamental reason we cannot have access to some 
of the {\it existing} degrees of freedom, which seems at odds with 
the dynamical setup. Rather, there should be a 
{\it dynamical} reason for why localizations in $\lambda_{00}$ 
seem FAPP out of reach. The idea of decoherence would be
that localizations in $\lambda_{00}$ are highly unstable against 
dynamical decoherence. 

% Regarding the charge superselection rule, one may ask the question 
% of what is the {\it quantum} physical r\^ole of the Coulomb
% field~\cite{Joos}. In some sense the analysis given here sheds 
% some light on this question. We have seen the necessity to 
% include the non-local canonical variables $\lambda_{lm},f_{lm}$ 
% and discussed their relation with the $\frac{1}{r^2}$ - part  
% of the electric field. Clearly, the arguments given here are 
% neither complete nor rigorous in any sense. What they suggest 
% is to relate the $\lambda_{lm}$  degrees of freedom to the 
% precise infrared structure of QED, for example along the lines of 
% \cite{Zwanziger1}\cite{Zwanziger2}\cite{Gervais-Zwanziger}
% (see below). Eventually this raises the question of how to 
% fully describe the state spaces of QED, which is known to be a
% notoriously difficult problem~(see e.g. \cite{Buchholz}). 

We have mainly focussed on the charge superselection operator 
$f_{00}$, although the foregoing considerations make it clear 
that by the same argument any two different asymptotic flux 
distributions also define different superselection sectors of 
the theory. Do we expect these additional superselection rules 
to be physically real? First note that 
for $l>0$ the $f_{lm}$ are not directly related to the multipole 
moments of the charge distributions, as the latter fall-off faster than 
$\frac{1}{r^2}$ and are hence not detectable on the sphere at 
infinity. Conversely, the higher multipole moments $f_{lm}$
are not measurable (in terms of electromagnetic fields) within any
finite region of space-time, unlike the charge, which is tight to 
massive particles; any finite sphere enclosing all sources has the 
same total flux. But the $f_{lm}$ can be related to the kinematical 
state of a particle through the retarded Coulomb field. In fact,
given a particle with constant momentum $\vec p$, charge $e$ and 
mass $m$, one obtains for the electric flux distribution at time 
$t$ on a sphere centered at the instantaneous (i.e. at time $t$)
%
%% BEGIN FOOTNOTE
%
particle position:\footnote{Formula (\ref{flux-inst}) 
requires a little more explanation: 
for a particle with general trajectory $\vec z(t)$ let 
$t'$ be the retarded time for the space-time point $(\vec x,t)$, 
i.e., $t'=t-\Vert\vec x-z(t')\Vert$ ($c=1$ in our units). 
Now we can use the 
well known formula for the retarded electric field 
(e.g. (14.14) in \cite{Jackson}) and compute the flux 
distribution on a sphere which lies in the space of 
constant time $t$, where it is centered at the retarded position 
$\vec z(t')$ of the particle. This flux distribution can be
expressed as function of the retarded momentum 
$\vec p':=\vec p(t')$ and the retarded direction 
$\vec n':=[\vec x-\vec z(t')]/\Vert\vec x-\vec z(t')\Vert$
as follows ($E':=\sqrt{{\vec p'}^2+m^2}$): 
\begin{equation}
{\varphi'}_{\vec p'}(\vec n')=
\frac{em^2}{4\pi}\frac{1}{[E'-\vec p'\cdot\vec n']^2}.
\label{flux-ret}
\end{equation}
If the particle moves with {\it constant} velocity 
$\vec v:=\dot{\vec z}$, the expression for the retarded Coulomb 
field can be rewritten in terms of the instantaneous position 
$\vec z(t)$ by using
$\vec z(t)=\vec z(t')+\vec v\Vert\vec x-\vec z(t')\Vert$. 
With respect to this center it is purely radial. 
Then one calculates the flux distribution on a sphere 
which again lies in the space of constant time $t$, but now 
centered at $\vec z(t)$ rather than $\vec z(t')$. This function 
can be expressed in terms of the instantaneous direction 
$\vec n:=[\vec x-\vec z(t)]/\Vert\vec x-\vec z(t)\Vert$ and the 
instantaneous momentum $\vec p:=\vec p(t)$. One obtains
(\ref{flux-inst}).}
%
%% END FOOTNOTE
%
\begin{equation}
\varphi_{\vec p}(\vec n)=\frac{em^2}{4\pi}
\frac{[\vec p^2+m^2]^{\frac{1}{2}}}
{[(\vec p\cdot\vec n)^2+m^2]^{\frac{3}{2}}}.
\label{flux-inst}
\end{equation} 
Hence different incoming momenta would induce different flux
distributions and therefore lie in different sectors. Given that 
these sectors exist this means that different incoming momenta 
cannot be coherently superposed and no incoming localized states 
be formed, unless one also adds the appropriate incoming infrared 
photons to just cancel the difference of the asymptotic flux 
distributions. This is achieved by imposing the `infrared coherence 
condition of 
%% BEGIN FOOTNOTE
Zwanziger~\cite{Zwanziger1}\footnote{Basically it says that  
the incoming scattering states should be eigenstates 
to the photon annihilation operators $a^{\rm in}_{\mu}(k)$ in the
zero-frequency limit.}
%% END FOOTNOTE
the effect of which is to `dress' the charged particles with infrared 
photons which just subtract their retarded Coulomb fields at
large spatial distances. Hence coherent superpositions of particles 
with different momenta can only be formed if they are dressed by 
the right amount of incoming infrared photons. 

As a technical aside we remark that this can be done without 
violating the Gupta-Bleuler transversality condition
$k^{\mu}a_{\mu}(k)\vert{\rm in}\rangle=0$ in the zero-frequency
limit, precisely because of the surface
term~(\ref{boundary})\cite{Gervais-Zwanziger}. This resolved an 
old issue concerning the compatibility of the infrared 
coherence condition on one hand, and the Gupta-Bleuler
transversality condition on the other~\cite{Haller,Zwanziger2}. 
From what we said earlier concerning the consistency of the 
variational principle in the 
presence of charged states, such an apparent clash of these two
conditions had to be expected: without the surface variables 
one cannot maintain gauge invariance at spatial infinity
(i.e. in the infrared limit) and at the same time include charged 
states. In the charged sectors the longitudinal infrared 
photons correspond to real physical degrees of freedom and it will
naturally lead to inconsistencies if one tries to eliminate them by
imposing the Gupta-Bleuler transversality condition also in the 
infrared limit. However, a gauge symmetry in the infrared limit 
can be maintained if one adds the asymptotic degrees of freedom 
in the form of surface terms. 

These remarks illustrate how the rich superselection structure
associated with different asymptotic flux distributions 
$f_{lm}$ renders the problem of characterizing state spaces in 
QED for charged sectors fairly complicated. 
This problem has been studied within various formalisms 
including algebraic QFT~\cite{Buchholz} and lattice
approximations, where the algebra of observables can be 
explicitly presented~\cite{Kijowski}. However, all this 
takes for granted the existence of the superselection rules, 
whereas we would like to see whether they really arise from 
some physical impossibility to localize the system in the
degrees of freedom labelled by $\lambda_{lm}$. What physics
should prevent us from forming incoming localized wave 
packets of charged {\it undressed} (in the sense above) 
particles, which would produce coherent superpositions of 
asymptotic flux distributions from the sectors with $l\geq 1$?
This cries out for a decoherence mechanism to provide a 
satisfying physical explanation. The case of charge 
superselection is, however, more elusive, since localizations in
$\lambda_{00}$ do not have an obvious physical
interpretation. Compare the controversy between
\cite{Aharonov,Mirman} on one side and \cite{WWW2}
on the other.

\vfill\eject

\end{document}